\documentclass[12pt,a4paper ]{article}
\usepackage{graphicx}

\newcommand{\frat}[2]{\frac{\textstyle #1}{\textstyle #2}}
\newcommand{\vf}[1]{\mbox{\boldmath $#1$}}
\newcommand{\nomer}[1]{\mbox{$\cal N$\hspace{-.5ex}\raisebox{.3ex}
           {\underline{\tiny 0}$\!$} #1}}

\begin{document}
 \begin{center}
{\Large \bf   Instanton Liquid at Finite Temperature and Chemical Potential of Quarks}\\
 \vspace{0.5cm} S.V. Molodtsov$^{1,3}$, G.M. Zinovjev$^2$\\
 \vspace{0.5cm} {\small $^1$Joint Institute for Nuclear Research, Dubna, 141980
 RUSSIA}\\
 \vspace{0.5cm} {\small $^2$Bogolyubov Institute for Theoretical Physics, ul. Metrolohichna 14-b, Kiev, 03680 UKRAINE}
\\ \vspace{0.5cm} {\small $^3$Institute of Theoretical and Experimental Physics, Moscow, 117259 RUSSIA}
\end{center}
\vspace{0.5cm}

\begin{center}
\begin{tabular}{p{16cm}}
{\small{Instanton liquid in heated and strongly interacting matter is studied using the variational principle.
The dependence of the instanton liquid density (gluon condensate) on the temperature and the quark chemical
potential is determined under the assumption that, at finite temperatures, the dominant contribution is given
by an ensemble of calorons. The respective one-loop effective quark Lagrangian is used.
}}
\end{tabular}
\end{center}
\vspace{0.5cm}

In current studies of strong-interacting matter under extreme
conditions, primary attention is focused on a description of its phase state at given temperature
and chemical potential. For definiteness, we consider that $T$ is the temperature of quarks and $\mu$
is the quark chemical potential (it is assumed that gluons are in thermodynamical equilibrium with quarks).
However, there is no approach making it possible to describe main features of the expected phase
diagram of quark-gluon matter at least qualitatively.

In the present study, we argue that the instanton liquid model of the QCD vacuum
 \cite{1} can shed light on some important features of a full picture. It is frequently noted that this
model offers a useful tool for obtaining phenomenologically plausible estimates in spite of the fact
that it is poorly justified because the typical size of an instanton is not properly fixed.
As of now, this fact is considered as inessential because a connection has been revealed between
limitations on the instanton size due to repulsion \cite{2}
and generation of mass of the gluon field in the framework of the quasi-classical approximation
 \cite{3}. The latter mechanism is a more general property of stochastic gluon fields than the former one.
We will discuss this question later. Here we assume that the problem of instanton size is solved
in one of the following scenarios: self-stabilization of the saturating ensemble
\cite{2},\cite{rez1},
freezing of the coupling constant \cite{rez2},
or influence of the confining component
\cite{rez3}. In the present study, primary attention is focused on a plausible qualitative model
describing a behavior of the gluon condensate.

In the beginning, we recollect the variational principle proposed in
 \cite{2} and the method of determination of the size of pseudoparticles and the density
of the instanton liquid and introduce notation for further considerations. In the model
of instanton liquid describing the QCD vacuum, it is assumed that the leading contribution
to the QCD generating functional is given by the background fields  representing superposition
of instantons in the singular gauge:
\begin{equation}
\label{1}
A^a_{\mu}(x;\gamma)=\frat2g~\omega^{ab}\bar\eta_{b\mu\nu}~a_\nu(y)~,~~~
a_\nu(y)= \frat{\rho^2}{y^2+\rho^2}~\frat{y_\nu}{y^2}~,~~~y=x-z~,~~\mu,\nu=1,2,3,4~.
\end{equation}
where $\rho$ is the size, $\omega$ is the matrix of color rotation, and $z$ is the position
of the center of a pseudoparticle (in the case of anti-instanton, the 't Hooft symbol
should be replaced as follows:
$\bar\eta \to \eta$). This being so, the QCD generating functional takes the form
\begin{equation}
\label{2}
 Y=\sum_{N=1}^\infty\frat{1}{N!}~\prod_{i=1}^N~\int d\gamma_i~d(\rho_i)~e^{-\beta~U_{int}(\gamma)}
 =\sum_{N=1}^\infty
\frat{1}{N!}~\prod_{i=1}^N~\int d\gamma_i~e^{-E(\gamma)}~,
\end{equation}
$$E(\gamma)=\beta~U_{int}(\gamma)-\sum~\ln d(\rho_i)~,$$
where
\begin{equation}
\label{drho}
d(\rho)=\frat{1}{\rho^5}~\widetilde\beta^{2N_c}~e^{-\beta(\rho)}~,
\end{equation}
is the instanton size distribution \cite{4};
 $d\gamma_i=dz_i~d\omega_i~d\rho_i$, and
$$\beta(\rho)=\frat{8\pi^2}{g^2}=-b~\ln (C_{N_c}^{1/b} \Lambda\rho)$$
is the action of a single instanton, where
($\Lambda=\Lambda_{\overline{MS}}=0.92 \Lambda_{P.V.}$) ё $C_{N_c}$,
depends on the renormalization scheme and, in the case under consideration, is given
by $C_{N_c}\approx\frat{4.66~\exp(-1.68 N_c)}
{\pi^2 (N_c-1)!(N_c-2)!}$, and $b=\frat{11~N_c-2~N_f}{3}$. We assume that
 $N_f$=2 here because the leading contribution to renormalization comes from
hard massless gluons and quarks.
The auxiliary function
$$\widetilde \beta=-b~\ln(\Lambda \bar\rho)~,$$
is evaluated at the scale $\bar\rho$ defined by an average size of pseudoparticles,
$U_{int}(\gamma)$  is considered assuming pair interaction dominance. Its contribution
has the form \cite{2}
$$\int d\omega_1~d\omega_2~dz_1~dz_2~U_{int}(\gamma_1,\gamma_2)=V~\xi^2~\rho_1^{2}~\rho_2^{2}~,$$
where $\xi^2=\frat{27~\pi^2}{4}\frat{N_c}{N_c^{2}-1}$.
The factor
 $\beta$ that appears in the exponent in formula
(\ref{2}) is also evaluated at the scale of an average size of pseudoparticles
 $\bar\rho$. Assuming that the instanton liquid is topologically neutral, we do not
introduce notation to distinguish between instantons and anti-instantons,
 $N$ denotes the ovarall number of pseudoparticles in volume $V$.

Since the interaction is independent of coordinates or orientation in color space,
it is natural to calculate the generating functional
 $Y$ on the basis of the effective one-particle distribution function $\mu(\rho)$,
which can be determined from the solution of the variational problem
\begin{equation}
\label{3}
 Y_1=\sum_{N=1}^\infty\frat{V^N}{N!}~\prod_{i=1}^N~\int ~d\rho_i~\mu(\rho)=
\sum_{N=1}^\infty\frat{V^N}{N!}~\prod_{i=1}^N~\int
d\gamma_i~e^{-E_1(\gamma)}~,
\end{equation}
$$E_1(\gamma)=-\sum~\ln \mu(\rho_i)~,$$
where the factor $V^N$ in (\ref{3}) is isolated
in order that the result be expressed in terms of the respective density
and convenience in interpretation of the function $\mu(\rho)$.
With regard to convexity of the exponential function, the generating functional
(\ref{2}) for every fixed $N$ partial contribution
can be estimated using the approximating inequality
\begin{equation}
\label{26} Y'\ge Y_a=Y'_1~\exp(-\langle E-E_1\rangle)~,
\end{equation}
where an average over approximate ensemble is implied.
In the case under consideration, the average of difference $\langle E-E_1\rangle$
is given by:
\begin{eqnarray}
\label{4} &&\langle E-E_1\rangle=\frat{1}{Y'_1}~\frat{1}{N!}
\int~\prod_{i=1}^N~ d \gamma_i~ [\beta~U_{int}-\sum\ln
d(\rho_i)+\sum \ln \mu(\rho_i)] ~e^{~\sum \ln \mu
(\rho_i)}=\nonumber\\ &&=\frat{N}{\mu_0}~\left(\int d \rho~ \mu
(\rho)~\ln \frat{\mu (\rho)}{d
(\rho)}+\frat{\beta}{2\mu_0}~\frat{N}{V}~\int d\rho_1
d\rho_2~\xi^2~\rho_1^{2}\rho_2^{2}~\mu (\rho_1) \mu
(\rho_2)\right)~,\nonumber
\end{eqnarray}
where $\mu_0=\int d \rho~\mu (\rho)$.

Variation of the functional  $\langle E-E_1\rangle$ with respect to $\mu(\rho)$ results
formally in the equation
$\mu(\rho)=e^{-1}~d(\rho)~e^{-n\beta\xi^2\overline{\rho^2}\rho^2}$
(where $n=N/V$ is the density of the instanton liquid). Here an unwanted factor of $e^{-1}$,
emerges. It can be excluded due to the fact that the approximate
functional $Y_a$ is independent of the constant factor of $C$ that can be added to the expressionfor $\mu(\rho)$.
For convenience, we set $C=e$, and therefore, arrive at
\begin{equation}
\label{5}
\mu(\rho)=d(\rho)~e^{-n\beta\xi^2\overline{\rho^2}\rho^2}~.
\end{equation}
Substituting this solution to the approximate functional, we obtain
$$Y_a=\frat{V^N~\mu_0^{N}}{N!}~e^{N\frac{\beta\xi^2}{2}(\overline{\rho^2})^2}~.$$
Defining suitable parameter $\nu$ the integral for determination $\mu_0$ can be represented in the form
\begin{equation}
\label{nu}
\mu_0=\Lambda^4~\int d\rho\Lambda~C_{N_c}\widetilde\beta^{2N_c}~(\rho\Lambda)^{b-5}~
e^{-\nu~\frac{\rho^2}{\overline{\rho^2}}}~.
\end{equation}
From the comparison of which with formula (\ref{5})  we obtain
\begin{equation}
\label{6}
\frat{\nu}{\overline{\rho^2}}=\beta \xi^2 n \overline{\rho^2}~.
\end{equation}
Provided that $\nu$ is known, this formula offers a relation
between the average instanton size and the density of the instanton liquid.
To find this relation, we consider the equation
$$\overline{\rho^2}=\frat{\int
d\rho~\rho^{b-3}~e^{-\nu~\frac{\rho^2}{\overline{\rho^2}}}} {\int
d\rho~\rho^{b-5}~e^{-\nu~\frac{\rho^2}{\overline{\rho^2}}}}=
\frat{{\overline{\rho^2}}~\nu^{-1}~\Gamma(\frac{b-4}{2}+1)}{\Gamma(\frac{b-4}{2})}~.$$
which gives $\nu=\frac{b-4}{2}$, and therefore,
$\mu_0=\Lambda^4~C_{N_c}\widetilde\beta^{2N_c}~\frat{(\rho\Lambda)^{2\nu}}{\nu^\nu}~
\frat{\Gamma(\nu)}{2}$. It should be noted that the factor of two in the denominator
of this expression stems from the integration measure $2\rho d\rho$, which, in its turn,
emerges in transformation to the Gaussian integral with respect to
$\rho$ squired. This factor was omited in
 \cite{2}; however, this fact has no noticeable consequences. The reason is that
the parameter $\Lambda$ is determined from a fit to some observable, for example,
to the pion decay constant. In so doing, everything is governed  by a choice of scale.
Moreover, it should be remembered that the instanton liquid model is merely a rough
approximation. From the above, we derive an approximate expression for the functional as
follows:
\begin{equation}
\label{7}
Y_a=\exp\left\{-N\left(\frat{\nu}{2}+1\right)~[\ln(n/\Lambda^4)-1]+
N\ln \left[C_{N_c}\widetilde\beta^{2N_c}(\beta\xi^2\nu)^{-\nu/2}\frat{\Gamma(\nu)}{2}\right]\right\}~.
\end{equation}
Now we find the value of $n$ at which the argument of the exponential approaches its maximum.
To do this, we should solve the equation
\begin{equation}
\label{8}
-\left(\frat{\nu}{2}+1\right)~\ln(n/\Lambda^4)+
\ln \left[C_{N_c}\widetilde\beta^{2N_c}(\beta\xi^2\nu)^{-\nu/2}\frat{\Gamma(\nu)}{2}\right]
+n~\frat{2N_c}{\widetilde\beta}\frat{d\widetilde\beta}{dn}-
n~\frat{\nu}{2\beta}\frat{d\beta}{dn}=0~.
\end{equation}

From the relation (\ref{6})  we obtain
$$\frat{1}{\beta}\frat{d\beta}{d\bar\rho}+\frat{1}{n}\frat{d n}{d\bar\rho}+\frat{4}{\bar\rho}=0~.$$
On the other hand,
$\frat{d\beta}{d\bar\rho}=-\frat{b}{\bar\rho}$, $\frat{d\widetilde\beta}{d\bar\rho}=
\frat{d\beta}{d\bar\rho}$.
We represent the derivative of  $\beta$ with respect to the density in the form
$\frat{d\beta}{dn}=\frat{d\beta}{d\bar\rho}/
\frat{d n}{d\bar\rho}$,
and obtain
\begin{equation}
\label{9}
\frat{d\beta}{dn}=\frat{1}{n}~\frat{b~\beta}{4\beta-b}~,
~~\frat{d\widetilde\beta}{d n}=\frat{d\beta}{dn}~.
\end{equation}
Thus we derive the expressionfor the instanton liquid density
\begin{equation}
\label{10}
n/\Lambda^4=\left[C_{N_c}\widetilde\beta^{2N_c}(\beta\xi^2\nu)^{-\nu/2}
\frat{\Gamma(\nu)}{2}\right]^{\frac{2}{\nu+2}}~
\exp\left[\left(\frac{4N_c}{\nu+2}\frac{\widetilde\beta}{\beta}-\frac{\nu}{\nu-2}\right)
\frac{\nu+2}{2\beta-\nu-2}\right]~.
\end{equation}
The contribution of the derivatives of the functions
 $\beta$ and $\widetilde\beta$ with respect to the density was disregarded in \cite{2}.
This contribution compensates for the above-mentioned factor of  $2$ though, as was noted above, this is not
essential. The obtained formula for the instanton liquid density by itself does not provide a solution
to the problem because it remains to solve the transcendental equation
 (\ref{6}) in $\bar\rho$, where the function $\beta$
  involves the logarithm of $\bar\rho$. To solve this equation, it is convenient to reformulate the problem
without resort to the explicit formula
 (\ref{10}) for the instanton liquid density.
By definition of the function
 $\beta$, the action of an isolated pseudoparticle must be positive.
This gives a limitation to the maximum size of an (anti-)instanton as follows:
$\bar\rho\Lambda C_{N_c}^{1/b}\leq 1$ (actually, $\bar\rho\Lambda \leq 1$).
Now we can solve the transcendental equation (\ref{8}),
by bisection of the segment. In so doing, a stationary value of
 $\bar\rho$ is determined at each step and the respective instanton liquid density
is determined from equation
 (\ref{6}). In the calculation of the generating functional, the contributions of the type
 $(\bar\rho\Lambda)^{2\nu}$ are used rather than the expression for the instanton liquid density.


Now we modify the variational principle in order to extend our description to the case of finite temperatures.
For this purpose, we employ calorons --- solutions of the Yang--Mills equations periodic in the Euclidean time.
The background field should be replaced by a superposition of calorons and anti-calorons as follows
 \cite{5}:
\begin{eqnarray}
\label{11}
A^{a}_\mu(x,\gamma)&=&-\frat{1}{g}~\omega^{ab}~\bar\eta_{b\mu\nu}~
\partial_\nu \ln \Pi,~\nonumber\\[-.2cm]
\\[-.25cm]
\Pi&=&1+\frat{\pi \rho^2 T}{r}\frat{\sinh(2\pi r T)} {\cosh(2\pi r
T)-\cos(2\pi\tau T)}~, \nonumber
\end{eqnarray}
where $T^{-1}$ is the period of the caloron,
 $r=|{\vf x}-{\vf z}|$ is the distance from the center of the caloron $z$ in three dimensional space, and
$\tau=x_4-z_4$ -- is the respective interval of "time". As the temperature tends to zero,
such solutions go over to (anti-)instantons in the singular gauge. Yet another modification
of the variational principle is the replacement of the distribution
 (\ref{drho}) in the instanton size by the function
\begin{equation}
\label{12}
d(\rho,T)=\frat{1}{\rho^{5}}~\widetilde \beta^{2 N_c} \exp [-\beta (\rho)-A_{N_c} T^2 \rho^2]~,
\end{equation}
where the coefficient  $A_{N_c}=\frat{1}{3}\left (\frat{11}{6} N_c-1\right)\pi^2$
accounts for the additional contribution to the action of each individual pseudoparticle.
It provides an approximation to a more exact expression
\begin{equation}
\label{13}
d(\rho,T)=d(\rho,0)~\exp \left\{-\left[
\frat{1}{2}~g^2T^2~\frat{(N_c+N_f/2)}{3}~\frat{4\pi^2\rho^2}{g^2}+
12~A(\pi\rho T)~[1+(N_c-N_f)/6]\right]\right\}~,
\end{equation}
constructed from the respective determinants \cite{6}. For our purposes it is sufficient to say
that the function $A(\pi\rho T)$ is determined by a shape of the pseudoparticle (\ref{11}).
This function was studied in the cited work; however, we do not use it in the present article.
It should be mentioned that the expansion up to the terms of the order $T^2$ can be used
as an approximate expression for the function $A(\pi\rho T)$
because, within the accuracy of the variational principle, only the terms up to order $\rho^2$
should be kept in the argument of the exponential in formula (\ref{12}).
The first term in formula (\ref{13}) is represented as a product of two factors; each
factor was interpreted in \cite{6}.
The first factor  is the square of the electric mass, that is, the temporal component of the gluon
polarization tensor evaluated at the zero energy and momentum. It has the form
\begin{equation}
\label{14}
m_{el}^{2}=\Pi_{44}(\omega=0,{\vf{p}}={\vf{0}})=g^2T^2~\frat{(N_c+N_f/2)}{3}~.
\end{equation}
The remaining components being equal to zero at zero energy-momentum.
Therefore, the magnetic mass vanishes. Note that the one-loop quark and gluon contributions
 to the polarization tensor are taken into account \cite{7},
the resulting sum being rearranged in order that the quark and gluon contributions in the medium
sum up to a finite value. This, formally, gives rise to a generation of the mass of the gluon field.
The second factor is the integral of the square of the fourth component $A_4$ of the field in
formula (\ref{11})
\begin{equation}
\label{15}
\int dy~ A^a_{4}(y)A^a_{4}(y)=\frat{4\pi^2\rho^2}{g^2}~.
\end{equation}
It is independent of the temperature \cite{8}.
It is seen that one can take into account only one-loop contribution
$\frat{1}{2}m_{el}^2~A^a_{4}A^a_{4}$ to the Lagrangian of the gluon field and neglect
other corrections. It was demonstrated  \cite{3} that the term $U_{int}$
describing the interaction of pseudoparticles can be brought in the form
 $\frat{1}{2}m^2~A^a_{\mu}A^a_{\mu}$, where
$m^2=9\pi^2~n~\bar\rho^2~\frat{N_c}{N_c^{2}-1}$.
Thus the interaction term also describes generation of the mass of the gluon field in the
instanton--anti-instanton medium in quasi-classical approximation. This being so,
chromoelectric and chromomagnetic fields are screened equally well provided that the instanton liquid
density is not equal to zero. It was shown that screening is a consequence of stochastic character
of the ensemble of gluon fields being unrelated to a specific instanton solution of the type
(\ref{1}) or details of the repulsion mechanism responsible for stabilization of the ensemble
\cite{2}. An application of these considerations to the (anti-)instanton solution
(\ref{1}) leads precisely to the formula for $U_{int}$.
It turns out that, in the caloron ensemble, screening of chromomagnetic fields and the
interaction term depends only weakly on the temperature. However, the anisotropy is negligible
small and the interaction term coincides with that obtained for the (anti-)instanton solution.
First it was found in  \cite{8}, where the instanton liquid was studied at non zero temperature.

The one-loop contribution of Plank gluons is proportional to
 $N_c$ (see formula (\ref{14})) and does not vary as the chemical potential becomes different from zero.
On the other hand, it is known that the one-loop fermion contribution in the medium can be calculated
exactly. It has no dangerous singularities \cite{9},
\cite{10}. The "temporal" component of the polarization tensor generated by a quark of definite flavor
has the form
\begin{eqnarray}
\label{16}
\Pi^{f}_{44}(k_4,\omega)&=&g^2~\frat{k^2}{\pi^2\omega^2}~\int_0^{\infty}
\frat{dp~p^2}{\varepsilon_p} ~n_p
\left[1+\frat{4\varepsilon_p^{2}-k^2}{8pk}\ln\frat{(k^2+2p\omega)^2+4\varepsilon_p^{2}k_4^{2}}
{(k^2-2p\omega)^2+4\varepsilon_p^{2}k_4^{2}}-\right.\nonumber\\
&-&\left.\frat{\varepsilon_pk_4}{p\omega}~\arctan \frat{8
p\omega~\varepsilon_p k_4}
{4\varepsilon_p^{2}k_4^{2}-4p^2\omega^2+k^4}\right]~,\nonumber
\end{eqnarray}
where $\omega=|{\vf{k}}|$, $k^2=\omega^2+k_4^{2}$, $\varepsilon_p=(m^2+{\vf{p}}^2)^{1/2}$, where $m$ --
is the quark mass, $n_p=n_p^{-}+n_p^{+}$, $n_p^{-}=(e^\frac{\varepsilon_p-\mu}{T}+1)^{-1}$,
$n_p^{+}=(e^\frac{\varepsilon_p+\mu}{T}+1)^{-1}$.
After summation over all components, the polarization tensor takes the form
\begin{equation}
\label{17}
\Pi^{f}(k_4,\omega)=g^2~\frat{2}{\pi^2}~\int_0^{\infty} \frat{dp~p^2}{\varepsilon_p}~ n_p
\left[1+\frat{2 m^2-k^2}{8pk}\ln\frat{(k^2+2p\omega)^2+4\varepsilon_p^{2}k_4^{2}}
{(k^2-2p\omega)^2+4\varepsilon_p^{2}k_4^{2}}\right]~.
\end{equation}
It is seen that, at $k_4=0$, and small values of $\omega$, the first term (that is, unit)
gives the dominant contribution to the gluon mass.
The spatial components are negligibly small. In particular, at  $\omega=0$ we obtain
\begin{equation}
\label{18}
\Pi^{f}(0,0)=\Pi^{f}_{44}(0,0)=g^2~\frat{2}{\pi^2}~\int_0^{\infty} \frat{dp~p^2}{\varepsilon_p}~n_p~,
\end{equation}
and at $T=0$ we arrive at
$\Pi^{f}(0,0)=g^2~\left[\frat{(\mu^2-m^2)^{1/2}\mu}{\pi^2}-
\frat{m^2}{\pi^2}\ln\frat{\mu+(\mu^2-m^2)^{1/2}}{m}\right]$.

The ultimate expression for the electric mass has the form
\begin{equation}
\label{19}
m_{el}^{2}=\left[g^2T^2~\frat{N_c}{3}+\sum_{f=1}^{N_f}\Pi^{f}(0,0)\right]~.
\end{equation}
In this approximation, the effect of the instanton liquid is completely accounted for by the quark mass
dynamically generated in the instanton medium. With such definition of mass, the formula
(\ref{14}) at $\mu=0$ and $T\neq 0$ should be modified.
The coefficient $\frat{1}{6}$ at $N_f$ should be replaced by $\frat{2}{\pi^2}$. However,
this replacement has only a little effect; self-consistency of our calculations will be discussed below.

 Using the integral (\ref{15}), which is also valid for the caloron solution,
we derive the expression for the distribution of pseudoparticles:
\begin{equation}
\label{20}
d(\rho;\mu,T)=d(\rho;0,0)~e^{-\eta^{2}(\mu,T)~\rho^2}~,~~\eta^2=2~\pi^2~
\left[T^2~\frat{N_c}{3}+\sum_{f=1}^{N_f}\Pi^{f}(0,0)\right]~.
\end{equation}
The one-loop quark contribution to the instanton action at zero temperature, finite
chemical potential, and $\omega\neq 0$ was
studied in detail in \cite{11} (see also \cite{12}, \cite{13}).
These studies make it possible to improve our description, however, we work within the approximation
(\ref{20}) and, moreover, we consider the limit of massless quarks.
A self-consistent calculation for the quark with dynamically generated mass can be the subject
of a separable study.

Necessary modifications in the variational principle are as follows. It was revealed that only the distribution
function $d(\rho;\mu,T)$ of pseudoparticles changes, whereas the repulsion interaction
 $U_{int}$ between pseudoparticles remains as before.
Similar to the case of instantons, we introduce the parameter $\nu$
satisfying the relation
\begin{equation}
\label{21}
\frat{\nu}{\overline{\rho^2}}=\eta^2+\beta \xi^2 n \overline{\rho^2}~,
\end{equation}
instead of (\ref{6}). Since the instanton liquid density is greater than zero, a new limitation
on the average size of pseudoparticle emerges
$\bar\rho\Lambda\leq\frat{\nu^{1/2}}{\eta}$.
If this limit is smaller than the limit descussed above, then it must be the starting point for
the determination of the equilibrium size of pseudoparticles by the bisection method.
The derivative of the function $\beta$ with respect to the density of the instanton liquid
can be determined from the relation (\ref{21}).
The result is
\begin{equation}
\label{22}
\frat{n}{\beta}\frat{d\beta}{dn}=\frat{b}{4\beta-b+\frac{2\eta^2\bar\rho^2\beta}{\nu-\eta^2\bar\rho^2}}~,
\end{equation}
it should be substituted in Eq. (\ref{8}) that determines the saddle-point.
The integral (\ref{18})  should be evaluated numerically because it cannot be calculated
analytically at arbitrary temperatures even though the quark mass equals zero. Thus we are ready to
determine the parameters of the instanton liquid everywhere over the $\mu$ -- $T$ plane.
For simplicity, the calculations are performed at zero quark masses. We neglect the light-quark
contribution to the respective determinants \cite{14} (see also \cite{15}).
We also disregard a possible temperature molecular behavior of instanton--anti-instanton pairs
 \cite{16}.

\begin{figure*}[!tbh]
\begin{center}
\includegraphics[width=0.5\textwidth]{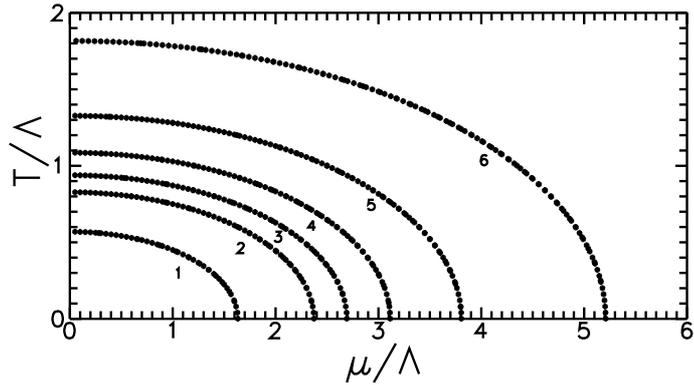}
\end{center}
  \vspace{-7mm}
 \caption{Lines of equal density of the instanton liquid in the temperature-chemical potential plane.
Curve $1$ corresponds to the density  $n=0.75~n_0$, where $n_0$ is the density at zero temperature
and chemical potential. Also shown are the densities from (curve $2$)  $n=0.5~n_0$ to (curve $6$)  $n=0.1~n_0$.
Curves $3$--$5$ correspond to intermediate densities at intervals of $0.1$.}
  \label{r}
\end{figure*}

The results of the calculations are shown in Fig. 1 by the lines of constant density. The instanton liquid density
is plotted in Fig 2. versus the temperature (at zero chemical potential) and versus the chemical potential
(at zero temperature). Though the conventional natation for the instanton liquid density at nonzero temperature
is $n=TN/V_3$, we use the label  $n$ which is more simple. At  $T\neq 0$,
and $\mu=0$, our results coincide with the results obtained in  \cite{8} and \cite{6}.
It sould be noted that our results are consistent with recent calculations on a lattice
at finite temperatures \cite{17}, \cite{18},
where a rapid decrease of the chromoelectric components in the respective correlation functions was
found. In our model, such suppression is due to the term
$\frat{1}{2}m_{el}^2~A^a_{4}A^a_{4}$
in the effective action; with neglect of this factor, the chromoelectric and chromomagnetic correlators
coincide. From this point of view, our calculations may seem inconsistent.
We use the caloron solution (\ref{11}),
which is symmetric under an interchange of chromoelectric and chromomagnetic fields.
However, the caloron components manifest themthelves in the observables differently
because of the anisotropy of the weight function.
In fact, our method of taking the gluon mass term into account is consistent only in perturbation
theory. In a complete study, one must find an analogue of the solution (\ref{11})
for the effective Lagrangian with the gluon mass generated for the chromoelectric field
and gain a self-consistent description of ensemble of pseudoparticles in the long-wave approximation
\cite{3}.

\begin{figure*}[!tbh]
\begin{center}
\includegraphics[width=0.5\textwidth]{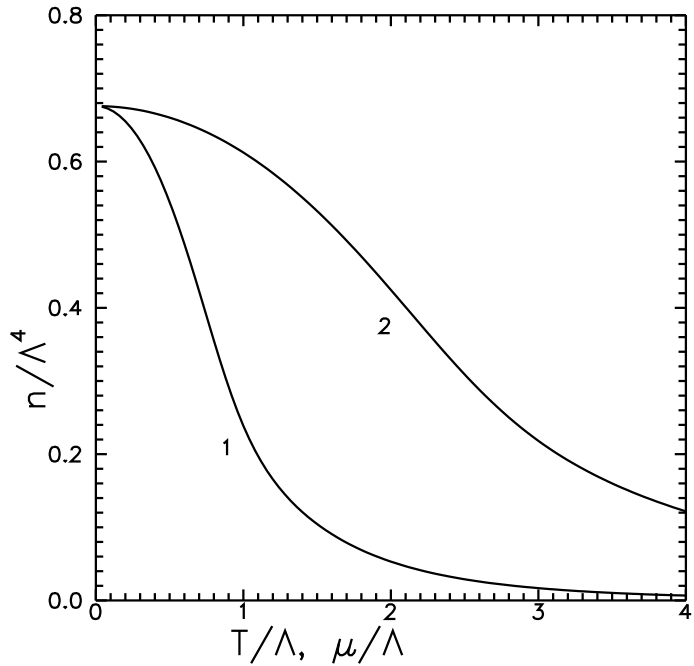}
\end{center}
  \vspace{-7mm}
 \caption{Instanton liquid density versus (curve 1) temperature and (curve 2) chemical
potential.}
  \label{rr}
\end{figure*}

It is of interest that the data on correlation functions for cooled configurations
\cite{17} are fitted well by the instanton ensemble  \cite{19}.
In so doing, the contribution of the terms of the second order in the instanton liquid density
($\sim n^2$) is in excellent agreement with the effect of the standard instanton ensemble with
the respective admixture of the perturbative component everywhere over the distance range chosen for a fit
 \cite{20}. This agreement indicates that the confining component is absent from the lattice configurations
isolated by cooling. It is surprising because lattice simulations with cooling were aimed at the searches for
a long-wave confining component. However, an interpretation of lattice simulations at finite temperature
presents difficulties because it is not clear what scale corresponds to the configarations used for
the measurements. The magnitude of deformation of the chromoelectric component of the solution for the effective
Lagrangian with the mass term is also poorly known. The scale of lattice configurations can, in principle,
be estimated using the scale at which the chromoelectric field decrease since only this scale has emerged in our
calculations.

In conclusion we note that, though we used only a rough approximation, the most important
features of the behavior of the instanton liquid density (gluon condensate) in the medium have
been revealed. The lines of equal density are markedly extended along the  $\mu$
axis because, according to the formula (\ref{19}), the most substantial gluon component
of screening vanishes at small temperatures. Typical values of
$T$ and $\mu$ at which the effects of the medium become significant are related to each other by
the formula
$\frat{(N_c+N_f/2)}{3}~(T/\Lambda)^2\sim
N_f~\frat{(\mu/\Lambda)^2}{\pi^2}\sim 1$,
which leads to a plausible coefficient of oblongness along the  $\mu$
axis $$\mu_c\sim
\sqrt{2\pi}~T_c~,$$ (at $N_c=3$ and $N_f=2$).
A fall in density evaluated with allowance for the dynamically generated quark mass
should begin at a greater value and be more steep. The reason is that, at chemical potentials less than
the quark mass, the quark contribution to screening is reduced.
This gives rise to formation of a plateau and concentration of the lines of equal density.
The dependence of the dynamical quark mass on the momentum
$\omega$ is significant at small temperatures leading to a decrease of screening approximately by a factor of two
\cite{11}.

We are grateful to A.E. Dorokhov for helful discussions.

This work was supported in part by grants STCU \#P015c, CERN-INTAS 2000-349, NATO 2000-PST.CLG
977482.


\end{document}